\documentclass[aps,prd,twocolumn,preprintnumbers,nofootinbib]{revtex4-1}
\usepackage{amsmath,amssymb,graphicx,natbib,bm}
\usepackage{enumerate}
\usepackage{hyperref}
\usepackage{breakurl}
\usepackage{color}

\newcommand{\GeV}{\ensuremath{\mathrm{GeV}}}

\usepackage[table]{xcolor}

\usepackage[latin1]{inputenc}

\begin{document} 

\preprint{PITT-PACC-1710}

\title{Thermal Dark Matter Through the Dirac Neutrino Portal}

\author{Brian Batell}
\affiliation{Pittsburgh Particle Physics, Astrophysics, and Cosmology Center, \\ 
Department of Physics and Astronomy, University of Pittsburgh, PA 15260, USA}

\author{Tao Han}
\affiliation{Pittsburgh Particle Physics, Astrophysics, and Cosmology Center, \\ 
Department of Physics and Astronomy, University of Pittsburgh, PA 15260, USA}

\author{David McKeen}
\affiliation{Pittsburgh Particle Physics, Astrophysics, and Cosmology Center, \\ 
Department of Physics and Astronomy, University of Pittsburgh, PA 15260, USA}

\author{Barmak Shams Es Haghi}
\affiliation{Pittsburgh Particle Physics, Astrophysics, and Cosmology Center, \\ 
Department of Physics and Astronomy, University of Pittsburgh, PA 15260, USA}

\begin{abstract}
We study a simple model of thermal dark matter annihilating to standard model neutrinos via the neutrino portal. A (pseudo-)Dirac sterile neutrino serves as a mediator between the visible and the dark sectors, while an approximate lepton number symmetry allows for a large neutrino Yukawa coupling and, in turn, efficient dark matter annihilation. The dark sector consists of two particles, a Dirac fermion and complex scalar, charged under a symmetry that ensures the stability of the dark matter.  A generic prediction of the model is a sterile neutrino with a large active-sterile mixing angle that decays primarily invisibly. We derive existing constraints and future projections from direct detection experiments, colliders, rare meson and tau decays, electroweak precision tests, and small scale structure observations. Along with these phenomenological tests, we investigate the consequences of perturbativity and scalar mass fine tuning on the model parameter space. A simple, conservative scheme to confront the various tests with the thermal relic target is outlined, and we demonstrate that much of the cosmologically-motivated parameter space is already constrained. We also identify new probes of this scenario such as multi-body kaon decays and Drell-Yan production of $W$ bosons at the LHC.

\end{abstract}

\date{\today}

\maketitle


\section{Introduction}
\label{sec:intro}
The search for non-gravitational dark matter (DM) interactions is one of the chief enterprises in modern experimental particle physics and observational astrophysics~\cite{Jungman:1995df,*Bergstrom:2000pn,*Bertone:2004pz,*Feng:2010gw}. While not required on general grounds, such interactions find strong motivation in the context of a cosmological origin of the dark matter abundance. Indeed, a compelling hypothesis is that DM is a thermal relic from the hot Big Bang, which requires non-gravitational couplings between DM and the Standard Model (SM) to ensure thermal contact and deplete the DM abundance. These couplings in turn predict a variety of novel phenomena associated with DM that can be sought through experiment and observation, offering the prospect of testing the thermal relic hypothesis. The most popular possibility for the non-gravitational couplings is the ordinary electroweak gauge interactions. This scenario is theoretically attractive due to the coincidence of the predicted and observed DM abundance for a TeV-mass relic with electroweak interactions and its potential connection to the Higgs naturalness problem, and has inspired an expansive and diverse experimental search program that has probed a significant portion of the parameter space. 

Nevertheless, thermal relic DM need not have SM gauge interactions. It is quite plausible that DM is instead a new electroweak gauge singlet particle, in which case an additional mediator particle is generally required to couple the dark and visible sectors~\cite{Boehm:2003hm,*Boehm:2003ha}. 
While there are in principle many ways that this mediation can occur, three renormalizable {\it portal} couplings stand out on account of their economy and uniqueness. 
These are the well-known vector portal~\cite{Okun:1982xi,*Galison:1983pa,*Holdom:1985ag}, Higgs portal~\cite{Silveira:1985rk,*Patt:2006fw}, and neutrino portal~\cite{Minkowski:1977sc,*Yanagida:1979as,*GellMann:1980vs,*Glashow:1979nm,*Mohapatra:1979ia,*Schechter:1980gr}:
\begin{eqnarray}
B_{\mu\nu} \, V^{\mu\nu} , ~~~~~  H^\dag H \, S, ~~~~~ \bar L H N +{\rm h.c.}, 
\end{eqnarray}
where $V_\mu$, $S$, and $N$ are new vector, scalar, and fermionic mediators, respectively, which can be straightforwardly coupled to the singlet DM particle. 
It is possible to distinguish two cases according to the dominant DM annihilation channel~\cite{Pospelov:2007mp}. The first case is {\it secluded annihilation}, 
in which dark matter annihilates directly to mediator particles. This occurs when DM is heavier than the mediator. The second, more predictive case, is {\it direct annihilation}, in which the dark matter annihilates through the mediator to SM particles. This occurs when DM is lighter than the mediator. In the secluded regime, the annihilation rate is set entirely by dark sector couplings. The mediator need only have a minuscule coupling to the SM to ensure kinetic equilibrium between the sectors, making it challenging to robustly test the scenario. In contrast, efficient direct annihilation necessitates a substantial mediator-SM coupling to avoid DM overproduction, leading to a predictive and experimentally testable scenario. For the predictive case of thermal DM directly annihilating to SM particles, a broad experimental effort is developing that will decisively test vector portal mediation~\cite{Alexander:2016aln}, while the case of Higgs portal mediation is already strongly constrained~\cite{Krnjaic:2015mbs}. 

Neutrino portal DM has been studied only a handful of times, despite sterile neutrinos $N$ being arguably the best motivated mediator candidate due to their role in neutrino mass generation. 
One likely reason for this is a theoretical prejudice for tiny neutrino Yukawa couplings, which in the simplest Type I seesaw with Majorana sterile neutrinos are generally far too small to allow efficient direct DM annihilation to SM neutrinos.
Most studies have therefore focused on the secluded regime, in which DM annihilates directly to the sterile neutrino mediators~\cite{Pospelov:2007mp,*Tang:2015coo,*Tang:2016sib,*Escudero:2016tzx,*Escudero:2016ksa,*Allahverdi:2016fvl,*Campos:2017odj,Batell:2017rol}. In this case the most robust signatures are indirect, and include cosmic gamma-rays antiprotons, and imprints of DM annihilation on the cosmic microwave background (see e.g., the recent study of Ref.~\cite{Batell:2017rol}).  

However, the case of direct annihilation to SM neutrinos can also be viable if the Yukawa couplings are large. Small neutrino masses are easily compatible with large neutrino Yukawa couplings provided sterile neutrinos are pseduo-Dirac states and an approximate $U(1)_L$ global lepton number symmetry is present.
A model incoporating these basic ingredients was first studied in Ref.~\cite{Bertoni:2014mva}, where the implications of a large neutrino-DM interaction for small scale structure were investigated. A similar model was analyzed in Ref.~\cite{Macias:2015cna,*Gonzalez-Macias:2016vxy}, where the focus was on heavy DM phenomenology. For other dark matter studies utilizing the neutrino portal, see Ref.~\cite{Falkowski:2009yz,*Falkowski:2011xh,*Cherry:2014xra}.

Our aim in this work is to provide a systematic analysis of thermal neutrino portal DM in the direct annihilation regime, over the entire cosmologically viable DM mass range from 1 MeV$-$1 TeV.
The basic model consists of a dark sector containing a Dirac fermion DM $\chi$ and a complex scalar $\phi$, along with a Dirac sterile neutrino mediator $N$, and is essentially the one constructed in Ref.~\cite{Bertoni:2014mva}. Beyond the particular motivation of Ref.~\cite{Bertoni:2014mva}, this model is of interest more generally as a scenario in which dark sector interactions with light neutrinos lead to thermalization and annihilation. We derive the existing constraints and future sensitivity projections from direct detection experiments, colliders, rare meson and tau decays, electroweak precision tests, cosmology. We also indicate the parameter regions that are favored by perturbativity and technical naturalness. In analogy with predictive vector and Higgs portal models, we also present a transparent strategy to compare these constraints and projections with the thermal relic target, under minimal and conservative assumptions on model parameters. We identify new probes of this model that can be undertaken with existing or near future experiments, including 3-body decays of stopped kaons and transverse momentum distributions of charged leptons in $W$ boson production at colliders. Furthermore, we emphasize the impact that a high statistics $\tau^+\tau^-$ sample at a future $B$-factory could have on testing this scenario.

The rest of this paper is organized as follows. In Sec.~\ref{sec:framework}, we outline the basic ingredients of the model, discuss its simple cosmology, and the decay modes of the new states. Section~\ref{sec:Npheno} describes constraints and probes of the heavy, mostly sterile neutrino in this scenario. The phenomenology of the DM in this model and ways to test it are in Sec.~\ref{sec:DMpheno}. In Sec.~\ref{sec:outlook} we conclude and discuss future prospects.

\section{Framework}
\label{sec:framework}
In this section we begin by defining a simple model in which DM couples to SM neutrinos through mixing generated with a neutrino portal coupling. We then discuss the cosmology and define the thermal relic target. We also discuss some basic features of the model that will be needed to understand the phenomenology, including the decays of the new heavy states, and the radiative couplings. We conclude this section with a brief discussion of technical naturalness in this scenario. 

\subsection{Model}
\label{sec:model}
As discussed in the introduction, the basic scenario we have in mind is thermal DM annihilating directly to SM neutrinos through the neutrino portal.\footnote{More precisely, the dark matter annihilates into very light, mostly SM flavor neutrinos which we refer to here as ``SM neutrinos'' for simplicity.} To allow for a large neutrino Yukawa coupling, and in turn an efficient annihilation rate, we take our mediator $N$ to be a Dirac particle. The dark sector is very minimal, and consists of a Dirac fermion DM candidate $\chi$ and a complex scalar $\phi$. 
Along with the kinetic terms, the Lagrangian is given by~\cite{Bertoni:2014mva}
\begin{align}
-{\cal L} & \supset  m_\phi^2 \, |\phi|^2  +  m_\chi \, \bar \chi \chi  + m_N \, \bar N N   \nonumber  \\
&  + \left[ \lambda_\ell \, \bar L_\ell {\hat H} N_R +\phi \, \bar\chi\left(y_L N_L+y_R N_R\right)+{\rm h.c.} \right],
\label{eq:Lagrangian}
\end{align}
where $H\ (\hat H = i\tau_2 H^*)$ is the Higgs doublet, $L_\ell=({\nu_\ell}_L,\ell_L)^T$ are SM lepton doublets with $\ell=e$, $\mu$, $\tau$ labeling the charged lepton mass eigenstates, $\lambda_\ell$ are the neutrino Yukawa couplings, and $y_{L, R}$ are couplings of the sterile neutrino mediator to the dark sector fields. The Lagrangian of Eq.~(\ref{eq:Lagrangian}) respects a global $U(1)_L$ lepton number symmetry, under which $L_\ell$, $N$ and $\phi^*$ have equal charges, as well as a global $U(1)_D$ dark matter number symmetry under which $\chi$ and $\phi$ have equal charges. At the level of Eq.~(\ref{eq:Lagrangian}), the lepton number symmetry forbids light SM neutrino masses, allowing us to take $\lambda_\ell$ as free parameters and in particular much larger than the usual naive Type I seesaw expectation. 

We will assume $m_\chi < m_\phi$, such that $U(1)_D$ symmetry ensures the stability of $\chi$. If, instead, $m_\chi > m_\phi$ then the $U(1)_D$ symmetry would render $\phi$ stable and it would be a good DM candidate. The phenomenology in this case would be essentially the same. The main difference is that the DM annihilation cross section is velocity-suppressed (in the limit that either $y_L$ or $y_R$ dominates), requiring larger couplings to provide efficient annihilation. Since the fermionic DM scenario can tolerate smaller couplings, it is more conservative. For this reason, we specify that $m_\chi < m_\phi$ and the fermion is the DM  for definiteness in our detailed study.

In the electroweak vacuum, $\langle H \rangle = v  = 174$ GeV, the SM neutrinos $\nu_i$ mix with $N$. 
Diagonalizing the Lagrangian, we find a heavy sterile Dirac neutrino state, which we label $\nu_4$. The  physical mass of this state is $m_4=\sqrt{m_N^2+\sum_\ell\lambda_\ell^2v^2}$, and
its left chiral component is a combination of sterile and active flavors,
\begin{align}
\nu_4=
\left(
\begin{array}{c}
U_{N4}^\ast N_L+\sum_\ell U_{\ell4}^\ast{\nu_\ell}_L \\
N_R
\end{array}
\right),
\end{align}
where the mixing angles are given by 
\begin{align}
\label{eq:angle}
U_{\ell 4}=\frac{\lambda_\ell v}{m_4}, ~~~~~~ \left|U_{N4}\right| = \frac{m_N}{m_4}=\sqrt{1-\sum_\ell\left|U_{\ell4}\right|^2}.
\end{align}
This is the crucial feature of the model: 
the mixing angles are not proportional to the light neutrino masses (in this limit, zero) and can be viewed as more or less free parameters.

The linear combinations of $N_L$ and ${\nu_\ell}_L$ that are orthogonal to ${\nu_4}_L$ are the light neutrinos, ${\nu_i}_L$ with $i=1$, $2$, $3$, which remain massless at this level. $N_L$ contains an admixture of light neutrinos, and so the light neutrinos interact with the DM via Eq.~(\ref{eq:Lagrangian}):
\begin{align}
\label{eq:Lagrangian2}
 &y_L \, \phi \, \bar\chi_R \, N_L+{\rm h.c.}        \\
& \to  y_L \left|U_{N4}\right|  \phi \, \bar\chi_R \, {\nu_4}_L - y_L \, \sqrt{1\!-\!\left|U_{N4}\right|^2} \, \phi \, \bar\chi_R \, {\nu_l}_L+{\rm h.c.} \nonumber
\end{align}
where $\nu_l$ is an admixture of light neutrinos $\nu_{1,2,3}$ with unit norm. The active neutrinos get an orthogonal admixture of heavy and light neutrinos,
\begin{align}
{\nu_\ell}_L= \sum_{i=1,2,3}U_{\ell i}{\nu_i}_L+U_{\ell 4}{\nu_4}_L
\end{align}
for $\ell=e$, $\mu$, $\tau$. 

The $4\times4$ matrix $U$ describes the relationship between gauge and mass eigenstates. Since the light neutrinos are all degenerate (i.e. massless) at the level we are dealing with them now, we can ignore mixing amongst the active neutrinos. Thus, the upper-left $3\times 3$ block of this matrix, corresponding to the usual Pontecorvo-Maki-Nakagawa-Sakata matrix~\cite{Pontecorvo:1967fh,*Pontecorvo:1957cp,*Maki:1962mu} and which governs ordinary neutrino oscillations, is unimportant for our purposes and the phenomenology is determined by the active-sterile mixing angles $U_{\ell 4}$.\footnote{See Sec.~\ref{sec:nu_osc}, however, for a discussion of atmospheric neutrino oscillations in the context of a large $\tau$-sterile mixing.}

There are a few simple ways to extend the model to incorporate light neutrino masses. 
The most interesting possibility is to endow $N$ with couplings that violate lepton number.
For example, one can add the Majorana mass term, $\mu \, \bar  N^c_R \, N_L +{\rm h.c.}$ as in the ``inverse seesaw'' scenario~\cite{Mohapatra:1986aw,*Mohapatra:1986bd}, or the Yukawa couplings, $\lambda'_\ell \, \bar L_\ell {\hat H}  N^c_R +{\rm h.c.}$ as done in~\cite{Malinsky:2005bi}.
This will lead to small neutrino masses, of order $\lambda_\ell^2 v^2 \mu/m_N^2$ or $\lambda'_\ell\lambda_\ell v^2/m_N$
respectively, governed by $U(1)_L$ breaking interactions, while the $\Delta L=0$ neutrino Yukawa coupling, $\lambda_\ell$, in Eq.~(\ref{eq:Lagrangian})
may still be large. These possibilities give $N$ the dual responsibility of generating neutrino masses and mediating DM interactions. In fact, only one Dirac sterile neutrino (made up of two Weyl fermions) is required to produce phenomenologically viable neutrino masses and mixings. Another possibility is that there are additional states $N'$ that participate in the mass generation, in either lepton-number--preserving or --violating ways with some other UV dynamics.
In any event, generating realistic neutrino masses while keeping the Yukawa couplings $\lambda_\ell$ large requires that $N$ is a Dirac or pseudo-Dirac state; in the latter case, the small mass splitting between the mass eigenstates will, however, have no significant phenomenological consequences for the parameter choices of interest to us.

Note that we have not written the renormalizable operator $\lambda_{\phi H} |\phi|^2 |H|^2$ which is allowed by the symmetries in Eq.~(\ref{eq:Lagrangian}) and is generated radiatively. After electroweak symmetry breaking, this operator contributes to the $\phi$ mass and to decays of the Higgs boson to $\phi\phi^\ast$ (which as we show in Sec.~\ref{sec:decays} are invisible at colliders). In a natural theory, we would expect the coefficient of this operator to be at least as large as the radiative contributions $\lambda_{\phi H}\gtrsim\delta_{\lambda_{\phi H}}\sim (y_L^2\lambda_\ell^2/16\pi^2)\log(\Lambda_{\rm UV}^2/m_4^2)$ where $\Lambda_{\rm UV}$ is a cutoff of the theory. However, given this estimate of the value of this operator, its effects on $m_\phi$ and the invisible Higgs width are subdominant to other contributions.

\subsection{Cosmology}
\label{sec:cosmo}
As is well-known, the relic density obtained through thermal freezeout is determined by the thermally averaged DM annihilation cross section $\langle  \sigma v \rangle$. The relevant process in this case is annihilation into light neutrinos, $\chi \bar \chi \rightarrow \nu \bar \nu$, and the cross section can be computed from the interaction Lagrangian, Eq.~(\ref{eq:Lagrangian2}):
\begin{align}
\label{eq:sigv}
\langle\sigma v\rangle&=\frac{y_L^4}{32\pi}\left(\sum_\ell\left|U_{\ell 4}\right|^2\right)^2\frac{m_\chi^2}{m_\phi^4}\left(1+\frac{m_\chi^2}{m_\phi^2}\right)^{-2}
\\
&\simeq 
1\,{\rm pb} \,
{\left(\! \frac{y_L\! \sqrt{\sum_\ell\left|U_{\ell 4}\right|^2}}{0.2} \right)}^{\!\!\! 4}\!
{\left(\!\frac{ 10\, \rm GeV}{m_\chi }\right)}^{\!\! 2}\!
{\left(\frac{3}{m_\phi/m_\chi}\right)}^{\!\! 4}, \nonumber 
\end{align}
where in the second step we have normalized the cross section to the canonical value $\langle\sigma v\rangle  = 3\! \times \! 10^{-26}$ cm$^3/$s $\simeq 1$ pb, which yields a DM abundance that is in close agreement with the measured value. 

\begin{figure*}
\begin{centering}
\includegraphics[width=0.99\linewidth]{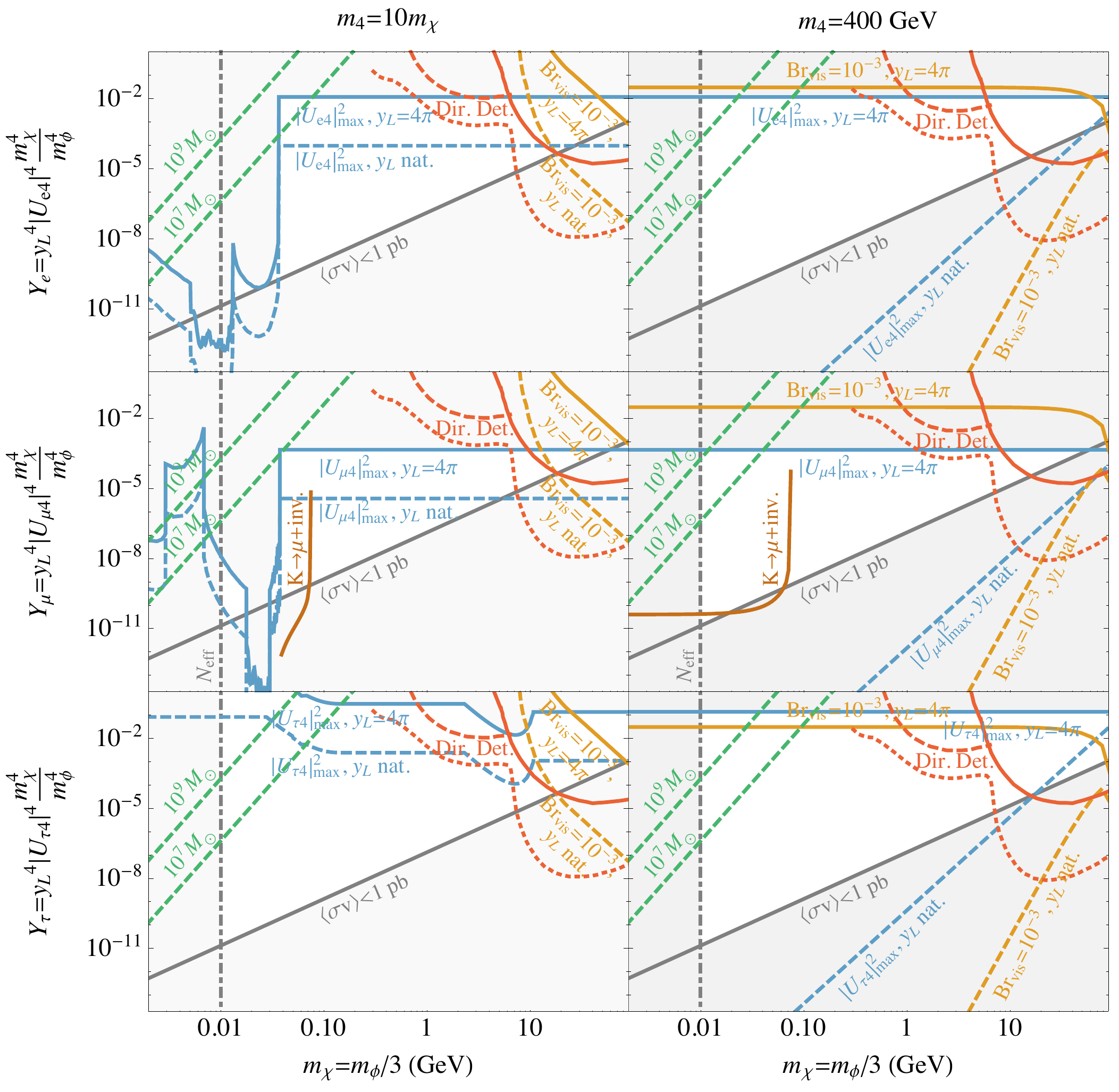}
\caption{Limits and constraints on the parameter $Y=y_L^4\left(\sum_i\left|U_{i4}\right|^2\right)^2\left(m_\chi/m_\phi\right)^4$ as a function of the DM mass $m_\chi$, assuming mixing dominantly with the $e$, $\mu$, $\tau$ flavors from top to bottom. We have fixed the mediator mass to be $m_\phi=3m_\chi$. On the left we take the heavy neutrino's mass to be $m_4=10m_\chi$ and on the right we fix it to $m_4=400~\rm GeV$. The blue lines show the limit on $Y$ given the upper limit on $\left|U_{i4}\right|^2$ (which are functions of $m_4$, see Fig.~\ref{fig:Ulimit}) and two different upper limits on $y_L$: 1) the perturbative limit $y_L<4\pi$ (solid) and 2) the fine tuning limit $y_L<4\pi m_\phi/m_4$ (dashed) where we require that radiative corrections to the $\phi$ mass are not larger than the $\phi$ mass itself and assume that $m_4$ is the cutoff scale. The  orange lines show where the heavy neutrino's visible branching ratio (through weak interactions) is $10^{-3}$ for the same two upper limits on $y_L$, pertubativity (solid) and fine-tuning (dashed); below these lines for either of these values of $y_L$ the visible branching ratio is smaller than $10^{-3}$. The solid gray line shows where the annihilation cross section is $1~\rm pb$, roughly the value required for a thermal relic. Above this line, the correct DM density can be easily explained by an initial asymmetry while below it requires a more intricate cosmological story. We also show direct detection limits through the effective DM coupling to the $Z$ (red, solid), as well as the direct detection cross sections corresponding the the ``$\nu$ floor'' (red, dotted). Future direct detection prospects are shown as red, dashed curves. Values of the coupling required to obtain small scale structure cutoff masses of $10^{7,9}M_\odot$, relevant for addressing the missing satellites problem are shown as dashed green lines. The ``$N_{\rm eff}$'' limit on the DM mass, $m_\chi\gtrsim 10~\rm MeV$, comes from CMB and BBN measurements of the effective number of relativistic degrees of freedom. See text for details.}
\label{fig:Yvsm}
\end{centering}
\end{figure*}

In analogy with simple vector portal DM scenarios~\cite{Izaguirre:2015yja,Alexander:2016aln}, it is useful to define a dimensionless parameter $Y$ that governs the annihilation rate, 
\begin{align}
Y&\equiv y_L^4\left(\sum_i\left|U_{i4}\right|^2\right)^2\frac{m_\chi^4}{m_\phi^4}.
\label{eq:Y}
\end{align}
In terms of this parameter Eq.~(\ref{eq:Y}), $\langle\sigma v\rangle \approx Y/(32 \pi m_\chi^2)$ and constraints on the model can be conveniently compared to the thermal relic target in the $m_\chi - Y$ plane.
In Fig.~\ref{fig:Yvsm} we show in this parameter space where the cross section Eq.~(\ref{eq:sigv}) is $3\times10^{-26}~\rm cm^3/\rm s=1~\rm pb$ as a function of DM mass $m_\chi$ with solid, dark gray diagonal lines. 
Along this line, the DM thermal relic density is close to the observed value. 
Above this line, the relic density is smaller than the measured density if DM is symmetric (equal populations of particles, $\chi$, and antiparticles, $\bar\chi$),
but it is easy to imagine obtaining the correct abundance in this region through an initial asymmetry (i.e., exactly what happens in the baryon abundance case). 
Below this line, if the DM is in thermal equilibrium then it naively overcloses the Universe, and obtaining the correct DM density here requires a more complicated nonthermal cosmology. 
From this point of view, parameter space above the $\langle\sigma v\rangle=1~\rm pb$ line is a well motivated target. Accordingly, we shade the region where $\langle\sigma v\rangle<1~\rm pb$ to indicate that it is disfavored in a simple thermal cosmology.
Our goal in this work is to place conservative bounds and projections on the cosmologically motivated region of parameter space, which suggests a particular set of benchmark model parameter choices.

We also display a variety of experimental constraints in Fig.~\ref{fig:Yvsm}, which will be surveyed in detail below. 
The first class of constraints, including direct detection experiments and the small scale structure of dark matter, scale in approximately the same way on $Y$, $m_\chi$ as the annihilation cross section and can be compared to the thermal target without further assumption on the model parameters. However, there are a number of other constraints on sterile neutrinos which limit the mixing angle as a function of the heavy neutrino mass and thus require further assumptions about the model parameters to obtain a constraint on $Y$. Noting Eq.~(\ref{eq:Y}), we see that for this class of constraints, the weakest (and thus most conservative) constraint on $Y$ as a function of $m_\chi$ is obtained by 1) saturating the experimental bound on the mixing angle, 2) taking $y_L$ close to nonperturbative values, and 3) maximizing the ratio of masses $m_\chi/m_\phi$ while keeping $m_\chi <m_\phi$. 
For these limits, we therefore fix a set of conservative benchmark values of $y_L = 4 \pi$ and $m_\chi/m_\phi =1/3$.

Note that to represent the constraints on the invisible heavy neutrino in the $m_\chi - Y$ plane, we further require an assumption about the heavy neutrino mass. 
There are two important comments in this regard, First, these constraints become stronger in the low $m_4$ regime. Thus to place conservative bounds we should take $m_4$ to be heavy. 
Second the dark scalar $\phi$ suffers from then usual naturalness problem associated with light scalar particles. The scalar receives a quadratically divergent correction to its mass coming from the same Yukawa coupling that enters into the DM annihilation cross section, $\delta m_\phi^2 \sim y_L^2 \Lambda_{\rm UV}^2/16 \pi^2$.
To conservatively implement a naturalness ``bound,'' we choose to cut this off at the scale $\Lambda_{\rm UV} = m_4$ (typically the heaviest particle in the low energy theory). Further UV contributions could be screened in  a low scale UV completion which stabilizes the $\phi$ mass, e.g. supersymmetry or compositeness. The operator $\lambda_{\phi H} |\phi|^2 |H|^2$ gives a contribution $\lambda_{\phi H} v^2$ to $m_\phi^2$. Given the radiative estimate of $\lambda_{\phi H}$ above, this contribution is suppressed by the relative factor $(|U_{\ell4}|^2 m_4^2/\Lambda_{\rm UV}^2)\log(\Lambda_{\rm UV}^2/m_4^2)$ so we do not include it in our ``bound.''

The naturalness ``constraint''  becomes weaker as $m_4$ is lowered. It is clear that the naturalness consideration is complementary to the invisible neutrino experimental constraints, since the latter prefer large $m_4$ while the former prefer small $m_4$. We thus show two sets of plots in Fig.~\ref{fig:Yvsm} with the following assumptions: 1) light $\nu_4$, fixing $m_\chi:m_\phi:m_4 = 1  :  3  :  10$, in the left column and 2) heavy $\nu_4$, fixing $m_4 = 400$ GeV, in the right column.

We also derive limits under the assumption that one of the mixing angles $|U_{e4}|$, $|U_{\mu4}|$, $|U_{\tau4}|$ dominates, while the others are negligible. This assumption is not necessarily conservative in terms of constraining the thermal target, since some of the limits may be weakened by ${\cal O}(1)$ factors while still obtaining a viable DM cosmology if two or more sizable mixing angles are present. This concern is relatively minor in our view, and is outweighed by the transparency of this assumption, which allows us to clearly display the flavor dependence inherent in certain limits. In this case we specify the flavor content of the coupling relevant for DM annihilation in Eq.~(\ref{eq:Y}) via $Y_\ell\equiv y_L^4|U_{\ell 4}|^4m_\chi^4/m_\phi^4$ for $\ell=e$, $\mu$, $\tau$.

Only the coupling $y_L$ enters the annihilation cross section Eq.~(\ref{eq:sigv}). The coupling $y_R$ does not lead to a tree-level DM-SM interaction, and for most DM masses plays no role in the cosmology. For simplicity, we therefore will assume in this work that $y_R$ is small and can be neglected.  However, we note that for $m_\chi \sim m_Z/2$ or $m_\chi \sim m_h/2$, DM can annihilate at one loop through an $s$-channel $Z$ or Higgs to the SM, which can potentially compete with the tree level process due to the resonant enhancement. It is easy to see that the radiative coupling of DM to the Higgs, $h \bar \chi \chi$, vanishes in the limit $y_R \rightarrow 0$.
As discussed further below, a radiative $Z \bar \chi \chi$ coupling, proportional to $y_L^2$, is still generated in this limit. For the benchmark choices described above, we have checked that annihilation through the $Z$ on resonance is subdominant to the tree level annihilation process Eq.~(\ref{eq:sigv}). It would be worthwhile to explore in more detail these resonance regions under different parameter assumptions, as this may allow certain constraints to be relaxed. However, these are fairly special regions of parameter space and their detailed exploration goes beyond our scope here. See Ref.~\cite{Macias:2015cna,*Gonzalez-Macias:2016vxy} for work in this direction. 

Light DM that is in thermal equilibrium with neutrinos can affect the number of relativistic degrees of freedom, $N_{\rm eff}$, as inferred by measurements of the cosmic microwave background (CMB)~\cite{Ade:2013zuv} and primordial light element abundances. The agreement of these with standard CMB and big bang nucleosynthesis (BBN) expectations can be used to set a lower limit on the DM mass of around $10~\rm MeV$~\cite{Serpico:2004nm,*Boehm:2013jpa,*Nollett:2014lwa,Iocco:2008va} which we show in Fig.~\ref{fig:Yvsm} with vertical dot-dashed lines.

\subsection{Decays of new states}
\label{sec:decays}
From the interactions in Eq.~(\ref{eq:Lagrangian2}) we can obtain the partial decay widths of the new unstable particles. The dark scalar $\phi$ decays to $\chi$ and a light anti-neutrino with a rate
\begin{align}
\Gamma_{\phi\to\chi \bar \nu_l}=\frac{y_L^2}{16\pi}\sum_i\left|U_{i4}\right|^2 m_\phi\left(1-\frac{m_\chi^2}{m_\phi^2}\right)^2.
\end{align}
The heavy neutrino $\nu_4$ decays (invisibly) to $\chi$ and $\phi$ through either $y_L$ or $y_R$. Assuming that $y_R$ is negligible, the rate for this is
\begin{equation}
\begin{aligned}
\Gamma_{\nu_4\to\chi\phi^\ast}& = 
\frac{y_L^2}{32\pi}\left|U_{N4}\right|^2m_4\left(1-\frac{m_\phi^2}{m_4^2}+\frac{m_\chi^2}{m_4^2}\right)
\\
&\quad\quad\times \lambda^{1/2}\left(1, \frac{m_\phi^2}{m_4^2},\frac{m_\chi^2}{m_4^2} \right),
\label{eq:decN-dark}
\end{aligned}
\end{equation}
where $\lambda(a,b,c)\equiv a^2+b^2+c^2-2ab-2ac-2bc$. This rate should be compared with the usual weak decay rate that occurs due to active-sterile mixing. 
The weak decay rate is negligible for $m_4 \ll m_W$, as it must proceed via off-shell weak bosons. For $m_4>m_W$, two body weak decays are kinematically allowed, and the rate for $\nu_4 \rightarrow \ell^- W^+$ is
\begin{align}
\Gamma_{\nu_4\to\ell^- W^+}&=\frac{G_F m_4^3}{8\pi\sqrt2}\left|U_{\ell 4}\right|^2\left(1-\frac{m_W^2}{m_4^2}\right)^2\left(1+\frac{2m_W^2}{m_4^2}\right)  \nonumber  \\
  &\simeq \frac{\lambda_\ell^2 m_4}{32 \pi},
  \label{eq:decN-weak}
\end{align}
where in the second step we have assumed $m_W \ll m_4$ and used the mixing angle relation Eq.~(\ref{eq:angle}). Comparing Eqs.~(\ref{eq:decN-dark}) and (\ref{eq:decN-weak}), we see that weak decays can be competitive if $\lambda_\ell$ is as large as $y_L$. 
However, given the mixing angle relation in Eq.~(\ref{eq:angle}), this only occurs in practice when $m_4$ is very large, of order 1 TeV or more. 
Therefore, weak decays of $\nu_4$ are subdominant over the entire parameter space consistent with thermal dark matter. To illustrate this, in Fig.~\ref{fig:Yvsm} we show contours where the weak branching fraction of the heavy neutrino is $10^{-3}$, fixing $y_L$ to either its perturbative maximum, $y_L=4\pi$ (orange, solid curves), or to its ``bound'' from naturalness, $y_L=4\pi m_\phi/m_4$ (orange, dashed curves). Below these lines, values of the weak branching fraction can be chosen to be less than $10^{-3}$, which is sufficient to reduce the limits derived from visible decays of the heavy neutrino (see, e.g.~\cite{Atre:2009rg}) to be subdominant. In this setup, where $N$ is Dirac or pseudo-Dirac, $\Delta L=2$ decays of the heavy neutrino are either absent or highly suppressed. In the two specific possibilities mentioned in Sec.~\ref{sec:model} where the light neutrino masses are due to either a lepton-number--violating Majorana mass $\mu$ or Yukawa coupling $\lambda^\prime_\ell$, the $\Delta L=2$ decay rate is suppressed relative to the $\Delta L=0$ rate by the factor $(\mu/m_N)^2$ or $(\lambda^\prime_\ell v/m_N)^2$, respectively.


\section{Sterile Neutrino Constraints}
\label{sec:Npheno}

As discussed in the previous section, the sterile neutrino mediator $\nu_4$ decays invisibly into the dark sector in our scenario. In Ref.~\cite{Bertoni:2014mva}, the limits on the mixing angle for invisibly decaying heavy neutrinos were presented (see also Ref.~\cite{deGouvea:2015euy}). These limits impact the cosmologically motivated DM parameter space since the annihilation cross section Eq.~(\ref{eq:sigv}) (or equivalently the $Y$ parameter in Eq.~(\ref{eq:Y})) depends on this mixing angle. In this section we briefly review the existing limits and take note of a few updates due to recent searches. The limits are summarized in the $m_4-|U_{i4}|^2$ plane in Fig.~\ref{fig:Ulimit}.

\subsection{$\mu,\tau$ decays}

A massive neutrino with non-vanishing $|U_{e4}|$ and/or $|U_{\mu4}|$ leads to a modified value of the Fermi constant extracted from the muon lifetime. Because the Fermi constant governs a host of precision electroweak and high-energy observables, the general agreement of this data~\cite{Tishchenko:2012ie,Olive:2016xmw} with the SM predictions constrains these mixing angles (labelled $\tau_\mu$/EWPT in Fig.~\ref{fig:Ulimit}). Similarly, the Fermi constant enters into the semi-leptonic weak decays used to measure the CKM elements, $V_{ud}$ and $V_{us}$, and CKM unitarity can therefore be used to derive limits on $|U_{\mu4}|$~\cite{Olive:2016xmw,Bertoni:2014mva}.
Furthermore, the lack of distortions in the $e^+$ energy spectrum in $\mu^+$ decays measured by the TWIST collaboration~\cite{Bayes:2010fg} can be used to constrain $|U_{e4}|$ and $|U_{\mu4}|$ for masses $m_4 < m_\mu-m_e$~\cite{Gninenko:2010pr,Bertoni:2014mva,deGouvea:2015euy}.

The constraints on mixing in the $\tau$ sector are generally much weaker than in the $e$ and $\mu$ sector. Relevant limits on $|U_{\tau 4}|$ can be derived from $\tau$ decays, including the leptonic decays $\tau\rightarrow e \bar \nu \nu$, $\tau\rightarrow \mu \bar \nu \nu$~\cite{Olive:2016xmw,Belous:2013dba,*Pich:2013lsa}, as well as certain hadronic modes like $\tau \rightarrow \nu 3 \pi$~\cite{Barate:1997zg,Bertoni:2014mva}.

\subsection{Rare meson decays}
 
 Rare meson decays, $M^+ \rightarrow \ell^+ \nu_4$, provide some of the best probes of an invisible heavy neutrino.
One striking signature is a peak in the energy spectrum of the outgoing lepton in the meson rest frame due to the two-body kinematics. Strong limits on $|U_{e4}|$ arise from peak searches in $\pi \rightarrow e \nu$~\cite{Britton:1992xv,*PIENU:2011aa,*Ito:2016nxh}, $K \rightarrow e \nu$~\cite{Yamazaki:1984sj}, and $B\rightarrow e \nu$~\cite{Park:2016gek} decays, while similar searches in the decays $\pi \rightarrow \mu \nu$~\cite{Abela:1981nf}, $K \rightarrow \mu \nu$~\cite{Hayano:1982wu,Artamonov:2014urb,Lazzeroni:2017fza}, and $B\rightarrow \mu \nu$~\cite{Park:2016gek} decays constrain $|U_{\mu4}|$.
Furthermore, a comparison of the experimental value of the ratio $\Gamma_{\pi \rightarrow e \nu}/\Gamma_{\pi\rightarrow \mu\nu}$ to its SM prediction can be used to set constraints on $|U_{e4}|$ and $|U_{\mu4}|$~\cite{Britton:1992pg,*Czapek:1993kc,*Marciano:1993sh}.

\subsection{Three body decays}
\label{sec:3body}
One of the most sensitive probes of sterile neutrinos comes from decays of a charged meson, $M^+$, to a charged lepton, $\ell^+$, and heavy, mostly sterile neutrino, as discussed in Sec.~\ref{sec:Npheno}. If the mass of the heavy neutrino, $m_4$, is larger than $m_M-m_\ell$, this two-body decay is kinematically forbidden from happening. However, in this model, if $m_\chi+m_\phi<m_M-m_\ell$, then three-body decays $M^+\to\ell^+\chi\phi$ can proceed through off-shell (heavy and light) neutrinos. In the limit that $m_4\gg m_M$, the rate for this decay is
\begin{align}
&\frac{1}{\Gamma_{M^+\to\ell^+\nu_\ell}}\frac{d\Gamma_{M^+\to\ell^+\phi\chi}}{dx}=\frac{y_L^2\left|U_{\ell 4}\right|^2}{32\pi^2}\left(1-\left|U_{\ell 4}\right|^2\right) \nonumber
\\
&\quad\frac{\sqrt{x^2-4x_\ell}}{x_\ell\left(1-x_\ell\right)^2}\frac{\left(1-x\right)x+2x_\ell}{\left(1+x_\ell-x\right)^3}\left(1+x_\ell-x-x_\phi+x_\chi\right) \nonumber
\\
&\quad\quad\times\lambda^{1/2}\left(1+x_\ell-x,x_\phi,x_\chi\right),
\label{eq:3body}
\end{align}
where $x=2E_\ell/m_M$ is the energy fraction carried by the charged lepton in the meson rest frame and $x_{\chi,\phi,\ell}=m_{\chi,\phi,\ell}^2/m_M^2$.

The signature of this decay is a charged lepton recoiling against something unobserved ($\chi$ and $\phi$ are both invisible) with a momentum different from the value expected for the standard decay into a charged lepton and massless neutrino, just as in the two-body case (where the decay products of the heavy neutrino are unobserved). The $\ell^+$ momentum is distributed over a range of values in the three-body case instead of being monochromatic as in the two-body case.

Armed with the expression in Eq.~(\ref{eq:3body}) (including terms that survive with finite $m_4$) for the decay rate, we can use the search for heavy neutrinos in $K^+\to\mu^++{\rm inv.}$ by the E949 experiment~\cite{Artamonov:2014urb} to set an upper limit on $y_L^2\left|U_{\mu 4}\right|^2$ or, equivalently, $Y_{\nu\mu}$. To perform this estimate we assume that the $\mu^+$ momentum distribution in the $K^+$ rest frame measured by E949 is well described by a power law background (dominantly from the radiative decay $K^+\to\mu^+\nu\gamma$) and ask what level of signal, $K^+\to\mu^+\chi\phi$, is allowed. We show the resulting 90\% CL upper limits on $Y_{\nu\mu}$ as a function of $m_\chi$ (assuming $m_\phi=3m_\chi$) for both the light $m_4$ and large $m_4$ cases in Fig.~\ref{fig:Yvsm}. In both cases, E949 data appear to be sensitive to an interesting region of parameter space close to thermal relic annihilation cross sections. The NA62 experiment plans to collect about $10^{13}$ kaon decays~\cite{Lazzeroni:2017fza,NA62:2017rwk} and could, assuming systematic errors can be kept under control, probe values of $Y_{\nu\mu}$ about an order of magnitude smaller than E949.

\subsection{Neutrino oscillations}
\label{sec:nu_osc}

Given the rather weak direct constraints on mixing in the $\tau$ sector, it is interesting to consider the consequences of mixing on neutrino oscillations. 
Notably, atmospheric neutrino oscillations are affected by $\nu_\tau - \nu_4$ mixing due to a suppression of the matter potential by the factor $(1-|U_{\tau_4}|^2)$ (see \cite{Bertoni:2014mva} for a detailed discussion). An analysis by Super-Kamiokande using the atmospheric muon neutrino zenith angle distribution leads to constraints on $|U_{\tau_4}|$~\cite{Abe:2014gda}. 

If the heavy neutrino is light, it can affect the number of relativistic species present during BBN. This sets a lower bound of $m_4\gtrsim 10~\rm MeV$~\cite{Iocco:2008va} which we show in Fig.~\ref{fig:Ulimit}, labelled ``$N_{\rm eff}$''.

\subsection{Invisible Higgs and $Z$ decays}

In addition to the flavor-specific constraints discussed above, invisible Higgs and $Z$ boson decays can be used to constrain the mixing angles. These are most relevant for larger values of $m_4$ where they become competitive with other constraints. We show the limit on the mixing angles from the constraint that the invisible branching of the Higgs is less than $0.24$ at $95\%$~C. L.~\cite{Khachatryan:2016whc} as solid blue lines in Fig.~\ref{fig:Ulimit}. We also show the reach that a future limit of $5\%$~\cite{Peskin:2013xra} on this branching could achieve as dotted blue lines. In setting these limits, we only consider decays of the Higgs to neutrinos through the Yukawa coupling with a rate proportional to $\lambda_\ell^2$. Since $\phi$ decays invisibly, $h\to\phi\phi^\ast$ decays generated by the operator $\lambda_{\phi H} |\phi|^2 |H|^2$ also contribute to the invisible Higgs width. However, this mode is suppressed relative to the neutrino mode by at least a further factor of $\lambda_\ell^2$ given the radiative estimate of the coefficient $\lambda_{\phi H}$ and we therefore ignore it.

The effect on the invisible branching of the $Z$ due to heavy sterile neutrinos that mix with the active neutrinos is well-known. The $90\%$~C. L. limits on the mixing angles that comes from the measurement of the invisible $Z$ width of $499.0\pm1.5~\rm MeV$~\cite{Olive:2016xmw} are shown in Fig.~\ref{fig:Ulimit} as solid, orange lines.

\subsection{LHC searches}

Furthermore, a heavy neutrino with a relatively large admixture of active flavors can be produced in large numbers in leptonic $W^\pm$ decays. Since, in the case we consider here, the heavy neutrino decays invisibly into the dark sector the only effect is a distortion of the kinematics of this decay. In particular, at a hadron collider, the $W$ transverse mass ($M_T^W$) endpoint or the transverse momentum ($p_T$) spectra of electrons and muons measured in Drell-Yan production of $W^\pm$ would be affected. 
For heavy neutrinos light enough to be produced on-shell in $W^\pm\to\ell^\pm$ decays, this would appear as a kink in the lepton $p_T$ at $p_T^{\rm peak}$ with the $M_T^W$ endpoint, $M_T^{\rm peak}$, shifted as
\begin{equation}
M_T^{\rm peak} = M_W \left(1-{m_4^2\over M_W^2}\right) , \quad 
p_T^{\rm peak} = {1\over 2} M_T^{\rm peak}.
\end{equation}
The relative size of the kink in this spectrum is at the level $\sim|U_{\ell 4}|^2$ for neutrinos kinematically allowed in $W$ decay. Neutrinos that are heavier than the $W$ simply dilute the rate by the factor $1-|U_{\ell 4}|^2$. Lepton $p_T$ spectra in $W^\pm$ production and decay at a hadron collider are very well studied because accurate measurements of these spectra, in particular their endpoints, are crucial in determining the $W^\pm$ mass. Since the recent ATLAS measurement of the $W^\pm$ mass~\cite{Aaboud:2017svj} has $e^\pm$ and $\mu^\pm$ $p_T$ spectra that agree with theoretical expectations at the subpercent level, we can reasonably expect a sensitivity to mixing angles of ${\cal O}(10^{-2})$ or smaller. This would be comparable to the limit from electroweak precision tests and therefore very interesting. 

To estimate the sensitivity of the measurement of lepton $p_T$ spectra at the LHC to $|U_{e4}|$ and $|U_{\mu 4}|$, we generated samples of $pp\to W^+\to e^+,\mu^++{\rm inv.}$ including massless and massive neutrinos~\cite{Alva:2014gxa,*Degrande:2016aje} corresponding to $4.1~\rm fb^{-1}$ of $7~\rm TeV$ $pp$ collisions using MadGraph5~\cite{Alwall:2014hca}, interfaced with Delphes~\cite{deFavereau:2013fsa} to model the detector response. We examine the resulting $e^\pm$ and $\mu^\pm$ $p_T$ distributions between $30~\rm GeV$ and $50~\rm GeV$ with bins of $0.5~\rm GeV$. Assuming that the standard model expectation describes these data and that statistical errors dominate (which are ${\cal O}({\rm few}\times 0.1\%)$ as in Ref.~\cite{Aaboud:2017svj}) aside from the overall normalization, we perform a fit to the underlying theory expectation (a function of $|U_{e4}|$ and $|U_{\mu 4}|$) at each value of $m_4$. We allow the overall measured cross section to vary by 2\% from the theoretical prediction to take the systematic error on the overall normalization, which comes mainly from the measurement of the total luminosity, into account. 

\begin{figure}
\begin{centering}
\includegraphics[width=1.15\linewidth]{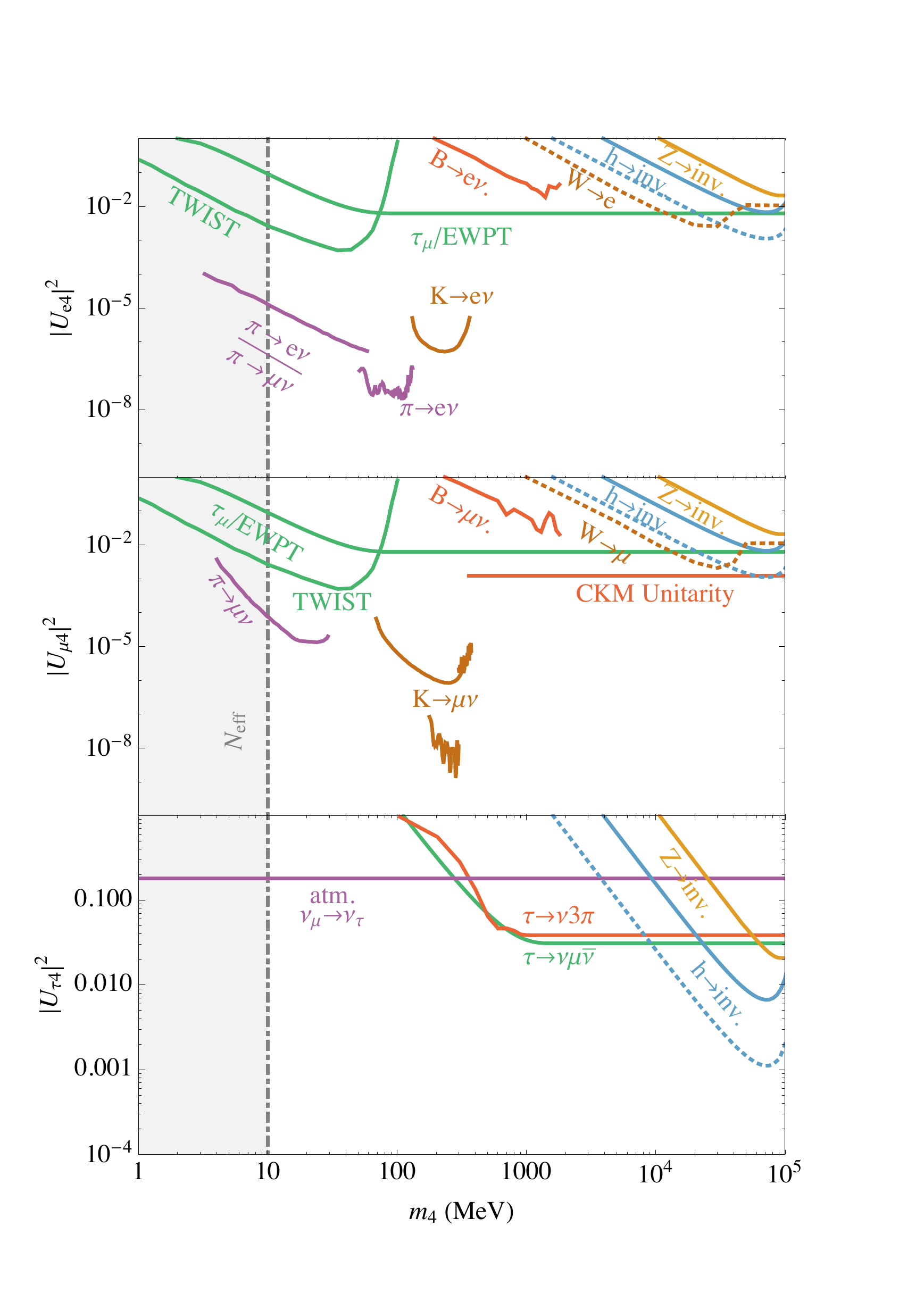}
\caption{90$\%$ C.L. upper limits on the mixing angles $\left|U_{\ell 4}\right|^2$ as functions of the heavy neutrino mass $m_4$ in the case that the heavy neutrino decays invisibly. Constraints on the mixing angle $|U_{e4}|$ (top) are obtained from searches in $\pi  \rightarrow e \nu$ and $K  \rightarrow e \nu$ decays, the ratio  $\Gamma_{\pi\rightarrow e\nu}/\Gamma_{\pi\rightarrow \mu\nu}$, electroweak precision tests (labelled $\tau_\mu/$EWPT), the lack of distortions in the $e^+$ spectrum in $\mu^+$ decays (labelled ``TWIST''), decays of $B$ mesons, and invisible $Z$ and Higgs decays.
Constraints on $|U_{\mu4}|$ (middle) come from peak searches in $\pi  \rightarrow \mu \nu$  and $K\rightarrow \mu \nu$ decays, electroweak precision and the $e^+$ spectrum in $\mu^+$ decays, decays of $B$-mesons, unitarity of the CKM matrix, and invisible $Z$ and Higgs decays. For $|U_{e4}|$ and $|U_{\mu 4}|$, we show our estimate of the reach from a search for massive, invisible neutrinos in Drell-Yan production of $W$ bosons at LHC, labelled ``$W\to e$'' and ``$W\to \mu$''.
Constraints on $|U_{\tau4}|$ (bottom) come from leptonic and hadronic $\tau$ decays, atmospheric neutrino oscillation studies, and invisible $Z$ and Higgs decays. The solid blue curves correspond to the limit ${\rm Br}_{h\to{\rm inv.}}<0.24$ while the dotted blue curves show the reach of a future limit ${\rm Br}_{h\to{\rm inv.}}<0.05$. This figure has been adapted and updated from Ref.~\cite{Bertoni:2014mva}. See text for further discussion and references.}
\label{fig:Ulimit}
\end{centering}
\end{figure}

In Fig.~\ref{fig:Ulimit} we show the resulting regions of parameter space ruled out at $90\%$ C.L.~with dashed brown lines labeled ``$W\to e$'' and ``$W\to \mu$''. As anticipated, the limit on the mixing angle for heavy neutrinos that can be produced on-shell is at the level of the statistical errors which dominate the shape of the $p_T$ spectra while for $m_4>m_W$ the limit corresponds to the uncertainty on the overall cross section. For lighter neutrino masses the difference in the lepton $p_T$ is difficult to distinguish from that for a massless neutrino so the limit is weakened.

\section{Dark Matter Phenomenology}
\label{sec:DMpheno}

\subsection{Direct Detection}
\label{sec:dirdet}
At one loop $\chi$ picks up an effective coupling to the $Z$ boson, which can be written after EWSB as 
\begin{align}
\label{eq:Z-coupling1}
 {\cal L} \supset \, a_Z \, \frac{g}{c_W}  \,Z_\mu \, \bar \chi \gamma^\mu P_R \chi, 
\end{align}
where the coupling $a_Z$ is given by
\begin{align}
\label{eq:Z-coupling2}
a_Z = |U_{N4}|^2(1-|U_{N4}|^2) \frac{y_L^2}{16 \pi^2} G\left(\frac{m_\phi^2}{m_4^2} \right) ,
\end{align}
with $G(x) = (x-1-\log{x})/(4(1-x)^2)$. In Eqs.~(\ref{eq:Z-coupling1}) and (\ref{eq:Z-coupling2}) we work in the limit $m_\chi \ll m_\phi, m_4$ and zero external momentum (we use the full loop integral for our numerical results). This coupling mediates spin-independent scattering of DM with nuclei vis $Z$ boson exchange, and therefore direct detection experiments provide an important probe of the model. The effective DM-nucleon spin-independent scattering cross section is given by 
\begin{align}
\sigma_n = \frac{\mu_n^2}{\pi}\frac{(Z f_p+(A-Z)f_n)^2}{A^2}
\end{align}
where $ f_n = G_F a_Z /\sqrt{2},\  f_p = -(1-4 s_W^2) G_F a_Z/\sqrt{2}$.

The strongest constraints in the high DM mass region come from recent results of XENON1T~\cite{Aprile:2017iyp} and PandaX-II~\cite{PandaXII} searching for spin-independent scattering of DM on nuclei, and are shown as a solid red curve in Fig.~\ref{fig:Yvsm}. These constraints rule out the thermal DM benchmark for DM masses heavier than about 15 GeV. We also show projections from SuperCDMS SNOLAB~\cite{Agnese:2016cpb}, which will cut further into $m_\chi \sim$ GeV mass region, as a dashed red curve. In addition, to get a sense of what region of parameter space can potentially be probed with direct detection, we show values of the coupling that correspond to a direct detection cross section at the ``neutrino floor''~\cite{Ruppin:2014bra} as a dotted red curve.

Clearly, direct detection experiments provide a powerful probe of the high DM mass region in this scenario, but it is important to note that these limits can be weakened or evaded altogether if the DM obtains a mass splitting. This can be implemented by softly breaking $U(1)_D$ to $Z_2$ by a Majorana mass terms for the chiral components of $\chi$. In this case upper limits on the DM mass are provided by constraints on the mixing angles for heavy, invisibly decaying neutrinos, notably electroweak precision tests as well as decays of electroweak bosons. Thus, these searches provide a complimentary probe of this scenario to direct detection.

There has been a significant effort devoted to exploring new methods of direct detection to probe low mass DM candidates~\cite{Battaglieri:2017aum}, although in our scenario the scattering cross section with electrons is unfortunately too small to be detected. 

\subsection{DM structure}
\label{sec:structure}
There has been growing evidence that observations of dark matter structure on subgalactic scales differs from the expectation from $N$-body simulations that assume that dark matter only interacts gravitationally (which, with respect to structure formation, essentially corresponds to the expectation for standard cold DM candidates). One of the longstanding problems in this area is the so-called ``missing satellites''~\cite{Klypin:1999uc,*Moore:1999gc,*Bullock:2010uy} problem---the apparent observation of fewer satellite galaxies of the Milky Way than expected. While the gravitational feedback of normal matter at small scales, resulting from complicated astrophysics, could resolve this discrepancy, it is also possible to address these issues with nongravitational interactions of dark matter. One such possibility is if the dark matter is relatively strongly coupled to the light neutrinos as in this scenario~\cite{Boehm:2000gq}.

A large coupling of dark matter to light (relativistic) neutrinos produces a pressure that that resists the gravitational collapse of the dark matter. Because both the number density of light neutrinos and the interaction cross section are larger at high temperatures, this pressure is important at early times when the Universe is hotter and eventually becomes unimportant at some critical temperature determined by the strength of the dark matter-neutrino interaction. Structures in the Universe form hierarchically, with smaller structures forming before larger ones, since only density perturbations with wavelengths smaller than the (expanding) horizon size can grow. The structures formed gravitationally by dark matter before matter-radiation equality provide the seeds for the growth of objects containing normal matter that we later observe. At early times when the dark matter-neutrino interactions are important, the pressure felt by the dark matter means that structures do not efficiently grow due to gravity, while, after the dark matter-neutrino interactions become unimportant, structures can form~\cite{Boehm:2001hm,*Boehm:2003xr}.

In other words, dark matter-neutrino interactions can address the missing satellites problem by suppressing the growth of small scale structures. The scale below which structures do not efficiently form in this scenario can be characterized in terms of a cutoff mass, $M_{\rm cut}$, which is the mass of dark matter inside the horizon at the critical temperature when DM-neutrino interactions become unimportant (for detailed discussion of these issues, see, e.g.,~\cite{Green:2005fa,*Loeb:2005pm}). This mass is a useful heuristic expected to correspond roughly to the size of the smallest gravitationally bound objects at late times. In this model the cutoff mass is estimated to be~\cite{Bertoni:2014mva}
\begin{equation}
\begin{aligned}
M_{\rm cut}\simeq 10^8M_\odot\left(\frac{g_\ast}{3.36}\right)^{-7/8}\left(\frac{0.1~\rm GeV}{m_\chi}\right)^{-14/4}Y^{3/4},
\end{aligned}
\label{eq:Mcut}
\end{equation}
assuming $m_\phi\gg m_\chi$. In this expression, $g_\ast$ refers to the effective number of relativistic degrees of freedom at the time of decoupling which we will always take to be $3.36$ which is relevant for the phenomenologically interesting case where the decoupling happens when the Universe's temperature is around a $\rm keV$.

We take values of the cutoff, $M_{\rm cut}$, between about $10^7$ and $10^9M_\odot$ to roughly represent DM-neutrino interactions strong enough to explain the missing satellites problem. An upper bound on $M_{\rm cut}$ of around $10^9M_\odot$ can be obtained analyzing Milky Way satellites~\cite{Strigari:2008ib,*Jethwa:2016gra}, looking for small scale structures in the Lyman-$\alpha$ forest~\cite{Viel:2013apy}, or through gravitational lensing of small, distant objects~\cite{Vegetti:2012mc,*Vegetti:2014lqa}.

\section{Summary and Outlook}
\label{sec:outlook}
We have presented a general analysis of thermal relic DM that annihilates directly to SM neutrinos through the neutrino portal. This possibility is very well motivated theoretically but has been relatively less well studied than other scenarios involving vector or scalar mediators.

The model described in Sec.~\ref{sec:model} is simple and economical, adding just three new states, fermions $N$ and $\chi$ and a scalar $\phi$, with masses from about $10~\rm MeV$ to ${\cal O}(100~\rm GeV)$. It allows for a very simple cosmological history, detailed in Sec.~\ref{sec:cosmo}, with the DM number density being set either by annihilation to light neutrinos or by an initial asymmetry, with annihilation reducing the number density to the measured level. The relatively large dark matter-light neutrino coupling needed for annihilation requires that the sterile neutrino mediator $N$ be (pseudo-)Dirac. We discussed the impact of radiative corrections to the scalar mass, identifying regions of parameter space that do not require fine-tuning.

In Sec.~\ref{sec:decays}, we pointed out an interesting  feature of this scenario: the fact that the heavy (mostly $N$) neutrino decays primarily invisibly into the dark sector, allowing for relatively larger active-sterile mixing angles. Sections~\ref{sec:Npheno} and \ref{sec:DMpheno} were devoted to fully exploring the model. We discussed and updated limits on the sterile neutrino in this scenario where it decays invisibly in Sec.~\ref{sec:Npheno}. 
The direct detection signature of this model was examined in Sec.~\ref{sec:dirdet} and we find that it sets an upper limit on the DM mass of around $10~\rm GeV$ in the simplest scenario. The impact on the small scale structure of DM, which could signal strong interactions between light neutrinos and DM, was also presented in Sec.~\ref{sec:structure}. Addressing this would likely imply a large mixing with the $\tau$ neutrino.

We identified several new measurements or analyses that can be done to probe large regions of viable parameter space in this setup.
First, we discussed the possibility that the heavy, invisibly decaying neutrino required for this scenario can affect the kinematics of charged leptons in Drell-Yan $W^\pm$ production at the LHC. The sizable active-sterile mixing angle needed as well as the large number of $W$ bosons produced at the LHC could allow for unconstrained regions of viable parameter space to be probed by such a measurement. In our simple mock analysis, we only considered $e^+$ and $\mu^+$ transverse momenta above $30~\GeV$ as in the ATLAS measurement of the $W$ mass~\cite{Aaboud:2017svj}. Including smaller $p_T$ values in this analysis would extend sensitivity to smaller values of the heavy neutrino mass $m_4$ and we urge any experimental analysis to push the lepton $p_T$ threshold as low as possible while still being able to deal with, e.g., issues from pile up.

Secondly, we discussed the sensitivity of three-body decays of kaons into the dark sector to this scenario. One virtue of this search is that it scales on the parameters of the model in the same way as the DM annihilation cross section and can therefore probe the parameter space without having to assume particular values of some parameters. We estimated the region of parameter space ruled out by the E949 experiment which collected around $10^{12}$ stopped kaon decays--this includes a large region of parameter space consistent with thermal relic DM that is unconstrained by other experiments. The NA62 experiment will collect around an order of magnitude more kaons and will therefore be able to probe even more of the viable thermal relic parameter space, providing a excellent test of this scenario.

Additionally, the active-sterile mixing angle is relatively less well constrained when it involves the $\tau$ flavor, particularly for $m_4\lesssim 300~\rm MeV$, which can allow for DM-neutrino interactions to be strong enough to affect small scale structure. In this region, the dominant constraint comes from the impact of nonstandard matter effects on the oscillation of atmospheric neutrinos as measured by Super-Kamiokande~\cite{Abe:2014gda}. It is important to note that this measurement is statistics limited and further data will probe a very interesting region, especially from the point of view of small scale structure effects. Lastly, the strongest limits on the mixing with the $\tau$ neutrino come from measurements of leptonic widths in $\tau^+\tau^-$ production at LEP~\cite{Abbiendi:1998cx,*Abreu:1999rb,*Acciarri:2001sg,*Abbiendi:2002jw,*Schael:2005am} and CLEO~\cite{Anastassov:1996tc} which involve samples of $10^5$ to $10^6$ $\tau^+\tau^-$ pairs. The upcoming Belle II experiment will collect roughly $4\times10^{10}$ $\tau^+\tau^-$ pairs with $50~\rm ab^{-1}$ of data. The possibility that the $\tau$ neutrino mixes with a sterile neutrino that decays into a dark sector provides strong motivation for studying $\tau$ decay rates and their kinematics with as much precision as possible. The extremely large data sample at Belle II could allow for much more parameter space to be probed or for a discovery to be made. We urge detailed experimental studies to be undertaken. Improving the reach on $|U_{\tau 4}|$ at collider experiments would test the possibility that DM-neutrino interactions affect small scale structure.

Models in which dark matter annihilates directly to light neutrinos are well-motivated, simple and far-reaching phenomenologically, in both particle physics and cosmology. New ideas to probe this class of models should be strongly encouraged. It could well herald a discovery that uncovers the particle nature of dark matter.

\subsubsection*{\bf Acknowledgements}
We thank Dorival Gon\c{c}alves and Satyanarayan Mukhopadhyay for helpful discussions. 
The work of BB and BSE is supported in part by the U.S. Department of Energy under grant No. DE-SC0015634, and in part by PITT PACC. 
The work of TH and BSE is supported in part by the Department of Energy under Grant No. DE-FG02-95ER40896, and in part by PITT PACC. 
The work of DM is supported by PITT PACC.

\bibliography{draft}

\end{document}